\documentclass[showpacs,prA,twocolumn]{revtex4}
\usepackage{amsmath,amssymb,amsthm}
\usepackage{epsfig}
\usepackage{graphicx}

\def\be#1{\begin{equation}\label{#1}}
\def\ee{\end{equation}}
\def\bea#1{\begin{eqnarray}\label{#1}}
\def\eea{\end{eqnarray}}
\def\sph{\hspace{.5em}}
\def\Eq#1{Eq.~(\ref{#1})}
\def\Fig#1{Fig.~\ref{#1}}

\def\no{\nonumber \\}
\def\tbf#1{\textbf{#1}}
\def\trm#1{\textrm{#1}}
\def\tit#1{\textit{#1}}

\def\mbf#1{\mbox{{\boldmath $#1$}}}
\def\half{\frac{1}{2}}

\def\expm2piOmega{e^{-2\pi\Omega}}

\def\sp{\hspace{.25em}}
\begin{document}

\title{Entanglement of Dirac fields in non-inertial frames}
\author{P. M. Alsing\footnote{alsing@hpc.unm.edu},
I. Fuentes-Schuller\footnote{Published before with maiden name Fuentes-Guridi}$^{\diamond}$,
R. B. Mann$^{\ddag *}$, and T. E. Tessier$^\dagger$}
\affiliation{
$^\ddag$Perimeter Institute, 31 Caroline Street North Waterloo, Ontario Canada N2L 2Y5\\
$^\diamond$ Instituto de Ciencias Nucleares, UNAM, A-postal 70-543 04510, Mexico D.F.\\
$^*$ Department of Physics, University of Waterloo, Waterloo, Ontario Canada N2L 3G1\\
$^\dagger$ Department of Physics and Astronomy, University of New Mexico,
Albuquerque, NM 87131-1156 }

\begin{abstract}
We analyze the entanglement between two modes of a free Dirac field
as seen by two relatively accelerated parties. The entanglement is
degraded by the Unruh effect and asymptotically reaches a
non-vanishing minimum value in the infinite acceleration limit. This
means that the state always remains entangled to a degree and can be
used in quantum information tasks, such as teleportation, between
parties in relative uniform acceleration. We analyze our results
from the point of view afforded by the phenomenon of entanglement
sharing and in terms of recent results in the area of multi-qubit
complementarity.
\end{abstract}

\maketitle

\section{Introduction}

Entanglement plays a central role in quantum information theory. It
is considered a resource for quantum communication and
teleportation, as well as for various computational tasks
\cite{ekertbook}. The importance of understanding entanglement in a
relativistic setting has received considerable attention recently
\cite{allthespecials,teleport,ivette,ball}. Such an understanding is
certainly relevant from a fundamental point of view, since
relativity is an indispensable component of any complete theoretical
model.  However,  it is also important in a number of practical
situations, for example, when considering the implementation of
quantum information processing tasks performed by observers in
arbitrary relative motion.

Entanglement was shown to be an invariant quantity for observers in
uniform relative motion in the sense that, although different
inertial observers may see these correlations distributed among
several degrees of freedom in different ways, the total amount of
entanglement is the same in all inertial frames
\cite{allthespecials}.  In non-inertial frames, entanglement was
first studied indirectly by investigating the fidelity of
teleportation between two parties in relative uniform acceleration
\cite{teleport}. More recently, the observer-dependent character of
entanglement was explicitly demonstrated by studying the
entanglement between two modes of a free scalar field as viewed by
two relatively accelerated observers \cite{ivette}.

A uniformly accelerated observer is unable to access information
about the whole of spacetime since, from his perspective, a
communication horizon appears. This can result in a loss of information
and a corresponding degradation of entanglement. In essence, the
acceleration of the observer effects a kind of ``enviromental
decoherence'', limiting the fidelity of certain quantum
information-theoretic processes.  A quantitative understanding of
such degradation in non-inertial frames is therefore required if one
wants to discuss the implementation of certain quantum information
processing tasks between accelerated partners.

In curved spacetime two nearby inertial observers are relatively
accelerated due to the geodesic deviation equation.  Accordingly,
the results of \cite{ivette} indicate that in curved spacetime even
two inertial observers will disagree on the degree of entanglement
in a given bipartite quantum state of some quantum field. Indeed, a
thorough investigation into entanglement in an expanding curved
spacetime shows that entanglement can encode information concerning
the underlying spacetime structure \cite{ball}.

In this paper we analyze the entanglement between two modes of a
Dirac field described by relatively accelerated parties in a flat
spacetime. We are interested in understanding how both the crucial
sign change in Fermi-Dirac versus Bose-Einstein distributions, and
the finite number of allowed states in fermionic systems due to the
Pauli exclusion principle (in contrast to the unbounded excitations
that can occur in bosonic systems), affect the degradation of
entanglement produced by the Unruh effect. We find that unlike the
bosonic case, where the entanglement degrades completely in the
infinite acceleration limit, in the fermionic case the entanglement
is never completely destroyed. We analyze the degradation of
entanglement in the system by applying the constraints of
entanglement sharing \cite{Coffman} and track the information
originally encoded in these quantum correlations by employing a set
of multi-qubit complementarity relations \cite{Tessier}. As in
\cite{ivette}, our results can be applied to the case that Alice
falls into a black hole whilst Rob barely escapes through eternal
uniform acceleration.

The remainder of this paper is organized as follows. In
Sec.~\ref{sec:setting} we consider two modes of a free Dirac field
that are maximally entangled from an inertial perspective. Two
parties, an inertial observer named Alice and a uniformly
accelerating observer named Rob, are each assumed to possess a
detector sensitive only to one of the two modes. Each measures the
field with his/her detector and the results are compared in order to
estimate the entanglement between the modes.

Section~\ref{sec:Unruh} discusses the Unruh effect for Dirac
particles as experienced by Rob. If a given Dirac mode is in the
vacuum state from an inertial perspective, then Rob's detector
perceives a Fermi-Dirac distribution of particles. This has a strong
effect on the entanglement that exists between Alice and Rob, and
therefore plays an important role in any quantum information task
they might perform that uses this entanglement as a resource.

In Sec.~\ref{sec:ent} we calculate the entanglement between the
modes from the perspectives of both Alice and Rob. Due to the
presence of a Rindler horizon, Rob is forced to trace over a
causally disconnected region of spacetime that he cannot access.
Accordingly, his description of the system takes the form of a
two-qubit mixed state. We calculate the entanglement using mixed
state entanglement measures such as the entanglement of formation
\cite{formation} and the logarithmic negativity \cite{negativity}.
We also estimate the total correlations (classical plus quantum) via
the mutual information \cite{mutual}. Our results show that the
entanglement of formation does not vanish as it does in the bosonic
case, but rather reaches a minimum of $1/\sqrt{2}$ in the limit that
Rob moves with infinite acceleration.

Since the fermionic system we are considering is accurately
described by a pure state of three qubits, we study the constraints
placed on the system by the phenomenon of entanglement sharing in
Sec.~\ref{sec:ent_sharing}.  Our analysis shows that no inherently
three-body correlations are generated in the quantum state. That is,
all of the entanglement in the system is in the form of bipartite
correlations, regardless of Rob's rate of acceleration.

Using complementarity relations applicable to an overall pure state
of three qubits, as well as to the various two-qubit marginals, we
identify the different types of information encoded in the quantum
state of our system in Sec.~\ref{sec:complementarity}.  This enables
us to study how specific subsystem properties depend on Rob's rate
of acceleration and to explain how some of the entanglement from the
inertial frame is able to survive in the fermionic system, even at
infinite acceleration. Finally, we summarize our results and suggest
possible directions for further research in
Sec.~\ref{sec:conclusions}.

\section{The setting}\label{sec:setting}

Consider a free Minkowski Dirac field in $3+1$ dimensions
\begin{equation}
i\gamma ^{\mu }\partial _{\mu }{\psi }-m\psi =0,  \notag
\end{equation}%
where $m$ is the particle mass, $\gamma^{\mu }$ the Dirac gamma
matrices, and $\psi $ is a spinor wavefunction. Minkowski coordinates
$x^{\mu} = (ct,\mathbf{x})$ with $\mu=\{0,1,2,3\}$ are the most suitable to describe the field from an
inertial perspective.
The field can be expanded in terms of the positive (fermions) and negative
(antifermions) energy solutions of the Dirac equation
$\psi _{k}^{+}$ and $\psi _{k}^{-}$ respectively, since they form a complete orthonormal set of modes,
\begin{equation}\label{Minkowski_field}
\psi =\int dk \, ( a_{k}\psi_{k}^{+}+b_{k}^{\dagger }\psi_{k}^{-}).
\end{equation}
In the above, $k$ is a notational shorthand for the wavevector $\mathbf{k}$ which
labels the modes.
The positive and negative energy Minkowski modes have the form
$$
\left(\psi_{k}^{\pm}\right)_{s} = \frac{1}{\sqrt{2\pi\omega_k}}
\,\phi^{\pm}_{s} \, e^{\pm i(\mathbf{k}\cdot\mathbf{x} - \omega_k t)},
$$
where $\omega_k = (m^2 + \mathbf{k}^2)^{1/2}$, and
$\phi_{s}=\phi_{s}(\mathbf{k})$ is a constant spinor with $s=\{\uparrow,\downarrow\}$ indicating
spin-up or spin-down along the quantization axis, satisfying the normalization relations \cite{mandl_shaw}
$\pm \bar{\phi}^{\pm}_{s}\,\phi^{\pm}_{s'} = (\omega_k/m)\,\delta_{ss'}$, $\bar{\phi}^{\pm}_{s'}\,\phi^{\mp}_{s'}=0$,
with the adjoint spinor given by $\bar{\phi}^{\pm}_{s} = \phi^{\pm\dagger}_{s}\,\gamma^0$.
The above positive and negative energy solutions satisfy the orthogonormality relations
$$
( \psi_{k}^{+}, \psi_{k'}^{+} ) = -(\psi_{k}^{-}, \psi_{k'}^{-} ) = \delta(k-k'), \quad
( \psi_{k}^{\pm}, \psi_{k'}^{\mp} ) = 0,
$$
where the Dirac inner product for two mode functions is given by
$$
\big( \phi(\mbf{x},t), \varphi(\mbf{x},t) \big)
= \int \, d\mbf{x} \, \phi^\dagger(\mathbf{x},t)\,\varphi(\mathbf{x},t).
$$
The modes $\psi_{k}^{\pm}$ are classified as positive and negative frequency with respect to
(the future-directed Minkowski Killing vector) $\partial_t$  for $\omega_k > 0$, i.e.
$$
\partial_t\, \psi_{k}^{\pm} = \mp\, i\,\omega_k \,\psi_{k}^{\pm}, \qquad \omega_k > 0.
$$

The operators $a_{k}^{\dagger },b_{k}^{\dagger}$ and $a_{k},b_{k}$
are the creation and annihilation operators for the positive and negative
energy solutions of momentum $k$ which satisfy the anticonmutation relations
\begin{equation}
\{a_{i},a_{j}^{\dagger }\}=\{b_{i},b_{j}^{\dagger }\}=\delta _{ij},  \notag
\end{equation}%
with all other anticommutators vanishing. The Minkowski vacuum state is
defined by the absence of any mode excitations in an inertial frame,
\begin{equation}
|0\rangle =\prod_{kk'}|0_{k}\rangle^{+}|0_{k'}\rangle^{-},  \notag
\end{equation}%
where the $\{+,-\}$ superscript on the kets is used to indicate the
particle and anti-particle vacua, respectively so that
$a_{k}|0_{k}\rangle^{+}=b_{k}|0_{k}\rangle^{-}=0$. We will use the
notation here, and throughout the rest of the work, that the mode
index ($k,k'\ldots$) will be a subscript affixed to the occupation
number inside the ket, and that the absence of a subscript on the
outside of the ket indicates a Minkowski Fock state. Since
$(a_{k}^{\dagger})^2=(b_{k}^{\dagger})^2=0$, there are only two
allowed states for each mode, $|0_{k}\rangle^{+}$ and
$|1_{k}\rangle^{+}=a_{k}^{\dagger}|0_{k}\rangle^{+}$ for particles,
and similarly for anti-particles.

Consider two maximally entangled fermionic modes in an inertial frame,
%\begin{equation}
%|\phi \rangle _{ar}=\frac{1}{\sqrt{2}}(|0\rangle _{k}^{+}|0\rangle_{r}^{+}
%+|1\rangle _{k}^{+}|1\rangle _{r}^{+}).  \label{eq:ent}
%\end{equation}
%%
\begin{equation}
|\phi_{k_A, k_R} \rangle
=\frac{1}{\sqrt{2}}(|0_{k_A}\rangle^{+}|0_{k_{R}}\rangle^{+} +
|1_{k_A}\rangle^{+}|1_{k_{R}}\rangle^{+}),  \label{eq:ent}
\end{equation}
where the subscripts
$A$ and $R$ indicate the modes associated with the observers Alice and Rob, respectively.
All other modes of the field are in the vacuum state, and
%
%where the subscripts
%$A$ and $R$ indicate the modes associated with the inertial
%observers Alice and Rob (initially at zero acceleration), respectively.
%All other modes of the field are in the vacuum state, and
%
therefore the state can be written as
$|\Phi\rangle=|\phi_{k_{A},k_{R}}\rangle\, \{\prod_{k\neq
k_{A},k_{R}} |0_k\rangle^{+}\prod_{k'}|0_{k'}\rangle^{-}\}$. Now
assume that Alice is stationary and has a detector sensitive only to
mode $k_{A}$. Rob moves with uniform acceleration and takes with him
a detector that only detects particles corresponding to mode
$k_{R}$. We ask the question of what is the entanglement between
modes $k_{A}$ and $k_{R}$ observed by Alice and Rob, given that Rob
undergoes uniform acceleration. Note that in order to determine the
amount of entanglement, Alice and Rob perform measurements which are
then compared by either party in order to estimate the correlations
in the results. Due to Rob's acceleration, at some point Alice's
signals will no longer reach Rob, but Rob's signals will always be
available to Alice, (see \Fig{fig1}). At this point only Alice can
compare the measurement results and estimate the entanglement of the
state. Let us now consider the state observed by Rob.

\section{Unruh effect for Dirac particles}\label{sec:Unruh}

Consider Rob to be uniformly accelerated in the $(t,z)$ plane
($c=1$). Rindler coordinates $(\tau,\zeta)$ are appropriate for
describing the viewpoint of an observer moving with uniform
acceleration. Two different sets of Rindler coordinates, which
differ from each other by an overall change in sign,
are necessary for covering Minkowski space \cite{note1}. These sets
of coordinates define two Rindler regions that are causally
disconnected from each other.
\begin{eqnarray}\label{Rindler_coords}
a t &=& \sph e^{a\zeta}\sinh(a\tau), \quad a z = \sph e^{a\zeta}\cosh(a\tau), \sph \trm{in region I}
\nonumber \\
&&\\ \notag a t &=& -e^{a\zeta}\sinh(a\tau), \quad a z =
-e^{a\zeta}\cosh(a\tau), \sp \trm{in region II}
\end{eqnarray}
where $a$ denotes Rob's proper acceleration. The above
set of coordinates both give rise to the same \tit{Rindler} metric
$$
ds^2 = dt^2 - dz^2 - d^2\mathbf{x}_{\perp} = e^{2a\zeta}\left(d^2\tau - d^2\zeta\right)
 - d^2\mathbf{x}_{\perp},
$$
where $\mathbf{x}_{\perp} = (x,y)$ are the same in both Minkowski and Rindler spacetimes.
\begin{figure}[h]
\begin{center}
\includegraphics[width=3.5in,height=2.5in]{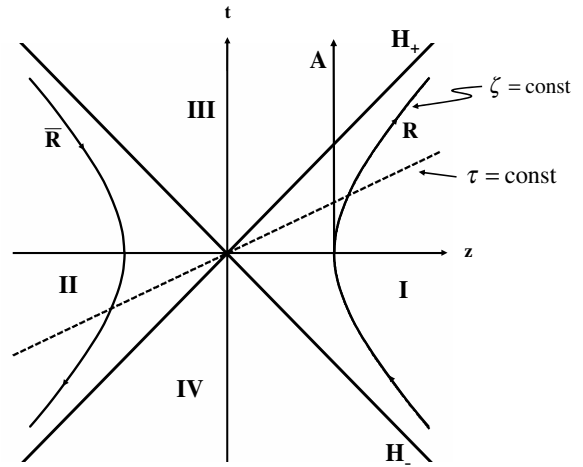}
\end{center}
\caption{Rindler spacetime diagram: Lines of constant position $\zeta$ are hyperbolas and lines of constant
proper time $\tau$ for the accelerated observer run through the origin. Note that while $\tau$ flows in the
direction of $t$ in region I, it flows in the direction of $-t$ in region II, (i.e the dashed line rotates
counter-clockwise for increasing values of $\tau$). A uniformly accelerated observer Rob ($R$) with
acceleration $a$ travels on a hyperbola constrained to region I, while a fictitious observer anti-Rob
$(\bar{R})$ travels on a corresponding hyperbola in region II given by the negative of Rob's coordinates.
The horizons $H_\pm$ are lines of $\tau=\pm\infty$ which
Alice ($A$) will cross at finite Minkowski times.}
\label{fig1}
\end{figure}
A particle undergoing eternal uniform acceleration remains
constrained to either Rindler region I or II and has no access to
the opposite region, since these two regions are causally
disconnected. Figure~\ref{fig1} serves to illustrate these ideas, as
well as to introduce our labeling scheme, where we refer to the
accelerated observer in region I as Rob (R) and to the corresponding
fictitious observer confined to region II (whose coordinates are the
negative of Rob's) as anti-Rob $\left(\rm{\bar{R}}\right)$.

The coordinates $(\tau,\zeta)$ have the ranges $-\infty < \tau,\,\zeta <\infty$ separately
in region I and in region II. This implies that region I and II each admit a separate
quantization procedure with corresponding positive and negative energy solutions
$\{\psi^{I+}_{k}, \psi^{I-}_{k}\}$ and $\{\psi^{II+}_{k}, \psi^{II-}_{k}\}$. Since the Rindler metric
is static (independent of $\tau$) it will admit solutions of the form
$e^{-i\omega\tau} \, \phi_{\alpha}(\zeta,\mathbf{x}_{\perp})$, with $\phi_{\alpha}$
a spatially dependent spinor \cite{note_on_spin}. Particles and anti-particles will be classified with
respect to the future-directed timelike Killing vector in each region. In region I
this is given by $\partial_{\tau}$ where
$$
\partial_{\tau} = \frac{\partial t}{\partial\tau}\,\partial_{t}
+ \frac{\partial z}{\partial\tau}\,\partial_{z}
=a\,(z\partial_{t} + t \partial_{z}),
$$
which is a boost into the instantaneous comoving frame of Rob. Thus, mode solutions
in region I having time dependence $\psi^{I+}_{k} \sim e^{-i\omega\tau}$ with $\omega>0$ represent
positive frequency solutions since $\partial_{\tau} \psi^{I+}_{k} = -i\omega \psi^{I+}_{k}$.
However, in region II $\partial_{\tau}$ points in the opposite direction of $\partial_{t}$
(increasing $\tau$ flows in the direction of $-t$, see \Fig{fig1}).
Hence in region II the future-directed
timelike Killing vector is given by $\partial_{-\tau} = -\partial_{\tau}$ \cite{BD,carroll}. Thus, a solution
in region II with time dependence $e^{-i\omega\tau}$ with $\omega>0$ is actually a negative
frequency mode since $\partial_{-\tau}\,e^{-i\omega\tau} = i\omega\,e^{-i\omega\tau}$. Hence, the
positive frequency mode in region II is given by $\psi^{II+}_{k} \sim e^{i\omega\tau}$ with $\omega>0$
satisfying $\partial_{-\tau} \psi^{II+}_{k} = -i\omega \psi^{II+}_{k}$.
Due to the causally disjoint nature of region I and II,
the modes $\psi^{I\pm}_{k}$ have support only in region I and vanish in region II, while the
opposite is true for the modes $\psi^{II\pm}_{k}$ in region II.
The Rindler modes satisfy orthonormality relations
analogous to the Minkowski modes \cite{note2} $(\psi^{\sigma\pm}_{k},\psi^{\sigma'\mp}_{k'})=0$, and
$(\psi^{\sigma\pm}_{k},\psi^{\sigma'\pm}_{k'}) = \pm\,\delta_{\sigma,\sigma'}\,\delta(k-k')$
where $\sigma\in\{I,II\}$.

In region I, let us denote $(c_{k}^{I},c^{I\dagger}_{k})$ as the annihilation
and creation operators for fermions (particles) and
$(d_{k}^{I},d^{I\dagger}_{k})$ as the annihilation and creation operators
for anti-fermions (anti-particles). The corresponding particle and
anti-particle operators in region II are denoted as
$(c_{k}^{II},c^{II\dagger}_{k})$ and $(d_{k}^{II},d^{II\dagger}_{k})$.
These obey the usual Dirac anti-commutation relations
$\{c_{k}^{\sigma},c^{\sigma'\dagger}_{k'}\} =$ $\{d_{k}^{\sigma},d^{\sigma'\dagger}_{k'}\} =$
%$\{c_{k}^{II},c^{II\dagger}_{k'}\} =$ $\{d_{k}^{II},d^{II\dagger}_{k'}\} =$
$\delta_{\sigma\sigma'}\,\delta_{kk'}$, with all other
anti-commutators, including those between operators in region I and
the causally disconnected region II, equaling zero.  Taking into
account the two sets of modes in each Rindler region, the Dirac
field can be expanded, in analogy to \Eq{Minkowski_field}, as
\begin{equation}\label{Rindler_field}
\psi = \int dk \left( c_{k}^{I}\psi _{k}^{I+}+d_{k}^{I\dagger }\psi _{k}^{I-}+c_{k}^{II}\psi
_{k}^{II+}+d_{k}^{II\dagger }\psi _{k}^{II-} \right).
\end{equation}

\Eq{Minkowski_field} and \Eq{Rindler_field} represent the
decomposition of the Dirac field in Minkowski and Rindler modes,
respectively. We can therefore relate the Minkowski and Rindler
creation and annihilation operators by taking appropriate inner
products. Using the Rindler orthogonality relations and
\Eq{Rindler_field} we have $c^{\sigma}_{k} =
(\psi^{\sigma+}_{k},\psi)$. Substituting \Eq{Minkowski_field} for
$\psi$ in this last expression yields \be{bogo_c} c^{\sigma}_{k} =
\int dk' \,\left( \alpha^{\sigma}_{kk'} \, a_{k'} +
\beta^{\sigma}_{kk'} \, b^{\dagger}_{k'}\right), \quad \sigma \in
\{I,II\} \ee where the Bogoliubov coefficients are given by the
inner product of the Rindler modes wavefunctions with the Minkowski
positive and negative frequency modes \be{bogo_coeffs}
\alpha^{\sigma}_{kk'} = \left( \psi^{\sigma+}_{k},
\psi^{+}_{k'}\right), \qquad \beta^{\sigma}_{kk'}  = \left(
\psi^{\sigma+}_{k}, \psi^{-}_{k'}\right). \ee A similar calculation
for $d^{\sigma}_{k}$ yields the corresponding expression \be{bogo_d}
d^{\sigma}_{k} = \int dk' \,\left( \alpha^{\sigma}_{kk'} \, b_{k'} +
\beta^{\sigma}_{kk'} \, a^{\dagger}_{k'}\right), \quad \sigma \in
\{I,II\} \ee with the same Bogoliubov coefficient as in
\Eq{bogo_coeffs}. In deriving \Eq{bogo_d} use has been made of the
following properties of the Dirac inner product: $(\phi_1,\phi_2)^*
= (\phi_1^*,\phi_2^* ) = (\phi_2,\phi_1)$. The calculation of the
Bogoliubov coefficients is straightforward, though lengthy and an
exercise in special functions. Details can be found elsewhere
\cite{takagi,rocio,mcmahon_alsing_embid}. For our purposes, the end
result of such calculations yields a relationship between the
Minkowski and Rindler creation and annihilation operators given by
the Bogoliubov transformation \bea{bogo_ops_a_bdag} \left[
\begin{array}{c}
  a_k \\
  b_{-k}^\dagger \\
\end{array}
\right] =
\left[
\begin{array}{cc}
  \cos r & -e^{-i\phi} \,\sin r \\
  e^{i\phi} \,\sin r & \cos r \\
\end{array}
\right] \,
\left[
\begin{array}{c}
  c^{I}_k \\
  d^{II\dagger}_{-k} \\
\end{array}
\right], \eea where $\tan r = \exp(-\pi\Omega)$ with $\Omega \equiv
\omega/(a/c)$, the ratio of the frequency $\omega$ to the only
naturally occurring frequency in the problem $a/c$, and $\phi$ is an
unimportant phase that can always be absorbed into the definition of
the operators. It is easy to see from \Eq{bogo_ops_a_bdag} and its
adjoint that given the anti-commutation relations of the Rindler
operators, the Minkowski anti-commutation relations are preserved.
In \Eq{bogo_ops_a_bdag} we have made the \tit{single mode
approximation} \cite{teleport}, which is valid if we consider Rob's
detector as sensitive to a single particle mode in region I such
that we can approximate the frequency $\omega_A$ observed by Alice
to be the same as the frequency $\omega_R$ as observed by Rob, i.e
$\omega_A \sim \omega_R = \omega$ \cite{note2.5}.
Note that this Bogoliubov transformation
mixes a particle in region I and an anti-particle in region II.
Correspondingly, the Bogoliubov transformation that mixes an
anti-particle mode in region I and a particle in region II is given
by \bea{bogo_ops_b_adag} \left[
\begin{array}{c}
  b_k \\
  a_{-k}^\dagger \\
\end{array}
\right] =
\left[
\begin{array}{cc}
  \cos r & e^{-i\phi} \,\sin r \\
  -e^{-i\phi} \,\sin r & \cos r \\
\end{array}
\right] \,
\left[
\begin{array}{c}
  d^{I}_k \\
  c^{II\dagger}_{-k} \\
\end{array}
\right]. \eea Since the anti-commutators between particle and
anti-particle operators and between region I and region II Rindler
operators are zero, it is easy to see that the Minkowski operators
in \Eq{bogo_ops_a_bdag} anti-commute with the Minkowski operators in
\Eq{bogo_ops_b_adag}, as they should (since $k$ and $-k$ represent
two separate modes).

As stated above, \Eq{bogo_ops_a_bdag} reveals that the Minkowski
particle annihilation operator is a Bogoliubov transformation
between a particle in region I of momentum $k$ and an anti-particle
in region II of momentum $-k$, i.e. a transformation that mixes
creation and annihilation operators. We can understand this in terms
of our previous discussion of the time dependence of positive
frequency Rindler modes in regions I and II. For a massless Dirac
field a positive frequency Rindler mode has the form
\cite{mcmahon_alsing_embid} $\psi^{I+}_k \sim \exp(i k\zeta -
i\omega\tau)$ (for the scalar case see \cite{carroll,takagi}) and a
positive frequency Rindler mode in region II has the form
$\psi^{II+}_k \sim \exp(i k\zeta + i\omega\tau)$. Thus, in order to
construct a positive frequency Minkowski mode $\psi^{+}$ that
extends $\psi^{I+}_k$ analytically from region I to region II, we
need a linear combination of $\psi^{I+}_k$ and $(\psi^{II+}_{-k})^*$
so that both have the space and time dependence $\exp(i k\zeta -
i\omega\tau)$. This means that the Minkowski operator $a_k$ must mix
$c^{I}_{k}$ and $d^{II\dagger}_{-k}$ as in \Eq{bogo_ops_a_bdag}.
Similarly, to construct a Minkowski mode that analytically extends
$\psi^{II+}_k$ from region II to region I we must form a linear
combination of $\psi^{II+}_k$ and $(\psi^{I-}_{-k})^*$ so that both
have the space and time dependence $\exp(i k\zeta + i\omega\tau)$.
This corresponds to $b_k$ mixing $d^{I}_{k}$ and
$c^{II\dagger}_{-k}$ as in \Eq{bogo_ops_b_adag}. For a massive Dirac
field the time dependence of the Rindler modes remains the same, but
the spatial dependence is more complicated
\cite{mcmahon_alsing_embid}. The above argument for the mixing of
Rindler particles (anti-particles) in region I (II) and
anti-particles (particles) in region II (I) goes through unchanged.

Having related Minkowski and Rindler creation and annihilation operators,
we now wish to relate the Minkowski vacuum to
the corresponding Rindler vacuum.
It is useful to note that \Eq{bogo_ops_a_bdag} can be written
as a two-mode squeezing transformation \cite{milburn_walls}
$$
\left[
\begin{array}{c}
  a_k \\
  b_{-k}^\dagger \\
\end{array}
\right] =
S\,
\left[
\begin{array}{c}
  c^{I}_k \\
  d^{II\dagger}_{-k} \\
\end{array}
\right]\,
S^\dagger
$$
for the single mode $k$ with $S$ given by
$$
S = \exp\left[r\,\left( c^{I\dagger}_k \,d^{II\dagger}_{-k} \,e^{-i\phi}
+  c^{I}_k \,d^{II}_{-k} \,e^{i\phi} \right)\right].
$$
Using the relation $e^A\,B\,e^{-A} = B + [B,A] + [B,[B,A]]/2! + [B,[B,[B,A]]]/3! + \cdots$
and the identity $[AB,C] = A\,\{B,C\} - \{A,C\}\,B$ we find
\bea{a_k}
a_k &=& S c^{I}_k S^\dagger = c^{I}_k - r e^{-i\phi}\,d^{II\dagger}_{-k}
- \frac{r^2}{2!}\,c^{I}_k + \frac{r^3}{3!}\,e^{-i\phi}\,d^{II\dagger}_{-k} + \cdots \no
 &=& \cos r \,c^{I}_k - e^{-i\phi} \sin r \,d^{II\dagger}_{-k}.
\eea
and
\bea{b_negk_dag}
b^\dagger_{-k} &=& S d^{II\dagger}_{-k} S^\dagger = d^{II\dagger}_{-k} + r e^{i\phi}\,c^{I}_{k}
- \frac{r^2}{2!}\,d^{II\dagger}_{-k} - \frac{r^3}{3!}\,e^{i\phi}\,c^{I}_{k} + \cdots \no
 &=& \cos r \,d^{II\dagger}_{-k} + e^{-i\phi} \sin r \,c^{I}_{k}.
\eea

Now $a_k$ and $b_{-k}$, respectively, annihilate the single
mode particle and anti-particle Minkowski vacua $a_k |0_k\rangle^+ =
0$ and $b_{-k} |0_{-k}\rangle^- = 0$. Since $a_k$ mixes particles in
region I and anti-particles in region II, we postulate that the
Minkowski particle vacuum for mode $k$ in terms of Rindler Fock
states is given by \be{vacM_1} |0_k\rangle^+ = \sum_{n=0}^1 A_n
\,|n_k\rangle^+_I\,|n_{-k}\rangle^-_{II}, \ee where \cite{note3}
\be{c_d_ops}
\begin{array}{lll}
  c^{I}_k |0_k\rangle^+_I = 0 & &d^{II}_{-k} |0_{-k}\rangle^-_{II} = 0, \\
   & & \\
  c^{I\dagger}_k |0_k\rangle^+_I = |1_k\rangle^+_I & &d^{II\dagger}_{-k} |0_{-k}\rangle^-_{II} = |1_{-k}\rangle^-_{II}. \\
\end{array}
\ee As a comment on notation, the Rindler region I and II Fock
states carry a subscript I or II, respectively, on the kets while
the Minkowski Fock states are indicated by the absence of an outside
subscript on the kets. Momentum mode labels are attached to the Fock
occupation number, and the $\{+,-\}$ ket superscript indicates a
particle or anti-particle state, respectively. Applying $a_k$ from
\Eq{a_k} to \Eq{vacM_1},  we have \bea{} 0 &=& a_k |0_k\rangle^+ \no
&=& \left( \cos r \,c^{I}_k - e^{-i\phi} \sin r \,d^{II\dagger}_{-k}
\right) \, \sum_{n=0}^1 A_n \,|n_k\rangle^+_I\,|n_{-k}\rangle^-_{II}
\no &=& (A_1 \,\cos r - A_0 \, e^{-i\phi}\,\sin
r)\,|0_k\rangle^+_I\,|1_{-k}\rangle^-_{II}\no &\Rightarrow& A_1 =
A_0 \,e^{-i\phi}\,\tan r. \eea Normalization $^+\langle 0_k|
0_k\rangle^+ =$ $|A_0|^2 + |A_1|^2=1$ yields $A_0 = \cos r$ so that
we finally arrive at
\bea{vacM} |0_k\rangle^+ &=& \cos r
\,|0_k\rangle^+_I\,|0_{-k}\rangle^-_{II} + e^{-i\phi}\,\sin r
\,|1_k\rangle^+_I\,|1_{-k}\rangle^-_{II}, \no
&=&\left[
\cos r \, + e^{-i\phi}\,\sin r \,c^{I\dagger}_{k}\,d^{II\dagger}_{-k}
\right] \,|0_k\rangle^+_I\,|0_{-k}\rangle^-_{II},
\eea
where the second equality in \Eq{vacM} is very useful for keeping track of
any possible transposition signs arising
from the anti-commuting operators when applying Minkowski operators to $|0_k\rangle^+$.
As an important example that we will use subsequently, a short calculation of
$a_k^\dagger \,|0_k\rangle^+$ using the adjoint of \Eq{a_k}
\be{a_k_dag}
a_k^\dagger = \cos r \,c^{I\dagger}_k - e^{+i\phi} \sin r \,d^{II}_{-k},
\ee
acting on the lower expression for
$|0_k\rangle^+$ in \Eq{vacM}, yields
\bea{1M}
a_k^\dagger \,|0_k\rangle^+ &=&
\left(
\cos^2 r \,c^{I\dagger}_{k} - \sin^2 r \,d^{II}_{-k}\,c^{I\dagger}_{k}\,d^{II\dagger}_{-k}
\right)\,\,|0_k\rangle^+_I\,|0_{-k}\rangle^-_{II}, \no
&=& \left(
\cos^2 r \,c^{I\dagger}_{k} + \sin^2 r \,c^{I\dagger}_{k}\,d^{II}_{-k}\,d^{II\dagger}_{-k},
\right)\,\,|0_k\rangle^+_I\,|0_{-k}\rangle^-_{II} \no
 &=& c^{I\dagger}_{k}\,|0_k\rangle^+_I\,|0_{-k}\rangle^-_{II}, \no
 |1_k\rangle^+ &=& |1_k\rangle^+_I\,|0_{-k}\rangle^-_{II}.
\eea
where in the second equality we have used the anti-commutation relations between
$c^{I\dagger}_{k}$ and $d^{II}_{-k}$ to obtain a minus sign upon transposition, and
in the third equality we have used the anti-commutation relations to write
$d^{II}_{-k}\,d^{II\dagger}_{-k} = 1 - d^{II\dagger}_{-k}\,d^{II}_{-k}$, noting
that the latter term annihilates the anti-particle vacuum. Lastly, one can easily verify
that $a_k^\dagger$ acting on $|1_k\rangle^+$ yields zero,
\be{a_k_dag_on_1_k}
a_k^\dagger |1_k\rangle^+ = \left( \cos r \,c^{I\dagger}_k - e^{+i\phi} \sin r \,d^{II}_{-k} \right) \,
|1_k\rangle^+_I\,|0_{-k}\rangle^-_{II} = 0,
\ee
ensuring that $(a_k^\dagger)^2 = 0$.

We note that the form of the Dirac particle vacuum for mode $k$ in
\Eq{vacM}, which can be written as $|0_k\rangle^+ = \cos
r\sum_{n=0}^1 \tan^n r \,|n_k\rangle^+_I\,|n_{-k}\rangle^-_{II}$, is
complementary to the form of the Minkowski charged scalar vacuum for
mode $k$, which is given by \cite{teleport,takagi}
$$|0_k\rangle^+ =
(\cosh r)^{-1} \sum_{n=0}^\infty \tanh^n r
\,|n_k\rangle^+_I\,|n_{-k}\rangle^-_{II}
$$
where in the latter case
$r$ is defined by $\tanh r = \exp(-\pi\Omega)$. Qualitatively, in
going from the scalar field to the Dirac field, scalar mode
functions are replaced by spinors, the infinite number of equally
spaced bosonic levels are replaced by two fermionic levels, and the
hyperbolic functions in the Bogoluibov transformation between
Minkowski and Rindler modes/operators are replaced by the
corresponding trigonometric functions (in essence $r\to i\,r$ in
going from the scalar to the Dirac field).

The two Minkowski states, $|0_k\rangle^+$ and  $|1_k\rangle^+$,
correspond to the particle field of mode $k$ observed by Alice.
On the other hand, an observer moving with uniform
acceleration $a$ in one of the regions has no access to field modes
in the causally disconnected region. Therefore, the observer must
trace over the inaccessible region, constituting an unavoidable loss
of information about the state, which essentially results in the
detection of a mixed state. Thus, when a Minkowski observer detects
a vacuum state $|0_k\rangle^+ \langle 0_k|$ for mode $k$, an
accelerated observer in region I sees a distribution of particles
according to the marginal state describing region I,
\bea{eq:modes}
\rho_k^I &=& Tr_{II}\left[|0_k\rangle^+ \langle 0_k|\right]\no &=&
\cos^2 r\,|0_k\rangle^+_I\,^+_I\langle 0_k| + \sin^2
r\,|1_k\rangle^+_I\,^+_I\langle 1_k|.
\eea
As the region I observer
accelerates through the Minkowski particle vacuum $|0_k\rangle^+$ of
mode $k$ his detector registers a number of particles given by
\bea{thermalization}
\lefteqn{^+\langle 0_k| \,c^{I\dagger}_k \,c^{I}_k|0_k\rangle^+
= Tr_{I,II}\left[c^{I\dagger}_k \,c^{I}_k\,|0_k\rangle^+ \langle 0_k|\right],} \no
&=& Tr_{I}\left[c^{I\dagger}_k\,c^{I}_k\,\rho_k^I\right] = \sin^2 r \;\; ^+_I\langle 1_k|c^{I\dagger}_k  \,c^{I}_k |1_k\rangle^+_I \no
&=& \sin^2 r = \frac{1}{e^{2\pi\Omega}+1} \equiv \frac{1}{e^{\hbar\omega/k_B T}+1}
\eea
where use has been made of
$\tan r = \exp(-\pi\Omega)$ with $\Omega = \omega c/a$, and we have
defined the Unruh temperature (where $k_B$ is Boltzmann's constant)
as \be{T_U}
 T =\frac{\hbar \,a}{k_B 2\pi c}.
\ee

Equation (\ref{thermalization}) is known as the Unruh effect
\cite{unruh} which shows that the uniformly accelerated observer in
region I detects a thermal Fermi-Dirac (FD) distribution of
particles as he traverses the Minkowski vacuum. Qualitatively, we
can understand the Unruh effect as follows. The constant proper
force $F$ that acts on a mass $m$ (a detector) in Rob's
instantaneous co-moving frame, to keep it under uniform
acceleration, can be written as $F=m a$ (which can be integrated as
$dp/d\tau = m a$ where $p = \gamma m v$, $\gamma =
(1-v^2/c^2)^{-1/2}$, to yield the hyperbolic orbits in
\Eq{Rindler_coords}). The work $\delta W$ that is performed on the
particle to keep it in uniform acceleration is the product of $F$
times a characteristic distance through which the force acts, which
we take to be the Compton wavelength of the particle $\delta z =
\hbar/(mc)$. This yields $\delta W = F\,\delta z = (m a) \,
\hbar/(mc) = \hbar \,a/c$, which is proportional to $k_B T$ in
\Eq{T_U}. Thus, the energy that is supplied to keep $m$ under
constant acceleration goes into exciting Rob's detector, and
curiously has a thermal spectrum (see \cite{alsing_milonni}). In the
general relativistic case this energy is supplied by the
gravitational field acting on a stationary observer (Rob) situated
outside a black hole, who experiences constant acceleration by
nature of his stationarity. A freely falling observer (Alice) who
eventually crosses the event horizon would experience the Kruskal
vacuum (the Minkowski-like vacuum of \Fig{fig1}) and thus detect no
Unruh radiation with a co-moving detector.

Note that in the case of a scalar field, \Eq{thermalization} has
a crucial minus sign in the denominator leading to a Bose-Einstein (BE) distribution of particles
appropriate for bosons. In both the scalar and Dirac case, the state of the Minkowski vacuum
for mode $k$ is a two-mode squeezed state (bosonic and fermionic, respectively), which
to a uniformly accelerated observer confined entirely to region I, is detected as a thermal state
(BE and FD, respectively).

In the following we investigate how the Unruh effect for Dirac
particles affects the entanglement between various Dirac modes.
Specifically, we study the entangled Bell state given by \Eq{eq:ent}
in the case that Rob is uniformly accelerated in region I. Assuming
a detector for Rob that is sensitive to a single Rindler particle
mode ($k\to k_R$), we decompose Rob's single particle Minkowski
states in \Eq{eq:ent} into the appropriate Rindler particle and
anti-particle states utilizing Eqs.~(\ref{vacM}) and (\ref{1M}). We
then proceed to evaluate various measures of entanglement. As stated
in the introduction, the advantage of utilizing a Dirac field over a
bosonic scalar field is that due to the finite occupation of the
fermionic states, we obtain finite dimensional density matrices that
lead to closed form expressions for the entanglement measures that
are more easily interpreted than their infinite dimensional bosonic counterparts.

%=================================================================================

%=================================================================================

\section{Fermionic entanglement from a non-inertial perspective}\label{sec:ent}

In this and subsequent sections we will use the following notation:
$A$ will indicate the inertial observer Alice, I will indicate the
uniformly accelerated observer Rob (R) confined to region I, and II
will indicate the fictitious complementary observer anti-Rob
($\rm{\bar{R}}$) in region II (see \Fig{fig1}), that arises from the
second (negative) set of Rindler coordinates in \Eq{Rindler_coords}.
Furthermore, since we are in the single mode approximation, we will
drop all labels ($k,-k,k_A,\ldots$) on states and density matrices
indicating the specific mode. Thus, the Minkowski particle mode
$|n_k\rangle^+$ for Alice will be written $|n\rangle_A$,  the
Rindler region I particle mode $|n_k\rangle^+_{I} \to
|n\rangle_{I}$, and the Rindler region II anti-particle mode
$|n_{-k}\rangle^-_{II} \to |n\rangle_{II}$ (for $n \in \{0,1\}$).
Likewise, we will refer to the ``Minkowski mode for Alice" simply as
``mode $A$", the Rindler particle mode in region I as ``mode I," and
the Rindler anti-particle mode in region II as ``mode II."

The density matrix for the Minkowski entangled state in
Eq.~(\ref{eq:ent}) is, from an inertial perspective
\begin{equation*}
\rho_{A,R}^{inertial}= \frac{1}{2}\left(
\begin{array}{cccc}
1 & 0 & 0 & 1 \\
0 & 0 & 0 & 0 \\
0 & 0 & 0 & 0 \\
1 & 0 & 0 & 1%
\end{array}%
\right),
\end{equation*}
in the basis $|00\rangle,|01\rangle,|10\rangle,|11\rangle$ with $%
|ab\rangle=|a\rangle_{A}\,|b\rangle_{R}$. Here, and only here, we
have used the subscript $R$ to indicate Rob's Minkowski Fock states
in \Eq{eq:ent}.

To describe the entanglement of the state as seen by an inertial
Alice and a uniformly accelerated Rob, we expand the Minkowski
particle states $|0\rangle_R$ and $|1\rangle_R$ into Rindler region
I and II particle and anti-particle states using Eqs.~(\ref{vacM})
and (\ref{1M}) to obtain
\begin{eqnarray*}
\rho_{A,I,II}&=&\frac{1}{2}
\left( \cos^2 r |000\rangle\langle 000| + \sin^2 r |011\rangle\langle 011| + |110\rangle\langle 110|\right) \\
&+& \frac{1}{2}
\left( \cos r \sin r |000\rangle\langle 011| + \cos r |000\rangle\langle 110| \right. \\
  &+& \hspace{1em}\left. \sin r |011\rangle\langle 110| + \trm{h.c.}\right),
\end{eqnarray*}
which we write in matrix form as
\begin{equation*}  \label{state}
\rho_{A,I,II}=\frac{1}{2}\left(
\begin{array}{cccccccc}
\cos ^{2}r & 0 & 0 & \cos {r}\sin {r} & 0 & 0 & \cos {%
r} & 0 \\
0 & 0 & 0 & 0 & 0 & 0 & 0 & 0 \\
0 & 0 & 0 & 0 & 0 & 0 & 0 & 0 \\
\cos {r}\sin {r} & 0 & 0 & \sin ^{2}{r} & 0 & 0 & \sin {%
r} & 0 \\
0 & 0 & 0 & 0 & 0 & 0 & 0 & 0 \\
0 & 0 & 0 & 0 & 0 & 0 & 0 & 0 \\
\cos {r} & 0 & 0 & \sin {r} & 0 & 0 & 1 & 0 \\
0 & 0 & 0 & 0 & 0 & 0 & 0 & 0%
\end{array}%
\right),
\end{equation*}%
in the basis $|000\rangle ,|001\rangle ,|010\rangle ,|011\rangle
,|100\rangle ,|101\rangle ,|110\rangle ,|111\rangle, $ where
for notational convenience we have defined
$|abc\rangle =|a\rangle_{A}\,|b\rangle_{I}\,|c\rangle_{II}$.
Note that the unimportant phase factor $\phi$ discussed in Section~\ref{sec:Unruh} has been absorbed
into the definition of the creation and annihilation operators.

Since Rob is causally disconnected from region II, we take the trace
over the mode in this region, which results in a mixed density
matrix between Alice and Rob
\begin{eqnarray}  \label{eq:state}
\rho_{A,I}=\frac{1}{2}\left(
\begin{array}{cccc}
\cos^{2}r & 0 & 0 & \cos{r} \\
0 & \sin^{2}r & 0 & 0 \\
0 & 0 & 0 & 0 \\
\cos{r} & 0 & 0 & 1%
\end{array}%
\right),
\end{eqnarray}
in the basis $|00\rangle,|01\rangle,|10\rangle,|11\rangle$ where $%
|ab\rangle=|a\rangle_{A}\,|b\rangle_{I}$.

To determine whether or not this state is entangled we use the
partial transpose criterion \cite{peres}. This criterion specifies a
necessary and sufficient condition for the existence of entanglement
in a mixed state of two qubits. If at least one eigenvalue of the
partial transpose of the density matrix is negative, then the
density matrix is entangled. The partial transpose is obtained by
interchanging Alice's qubits
$(|a_A\,b_I\rangle\langle c_A\, d_I | \to |c_A\,b_I\rangle\langle a_A\, d_I |)$
\begin{equation*}
\rho_{A,I}^{T}=\frac{1}{2}\left(
\begin{array}{cccc}
\cos ^{2}r & 0 & 0 & 0 \\
0 & \sin ^{2}r & \cos {r} & 0 \\
0 & \cos {r} & 0 & 0 \\
0 & 0 & 0 & 1%
\end{array}%
\right) .
\end{equation*}%
$\rho_{A,I}^{T}$ has eigenvalues $(1,1,\cos^2 r, -\cos^2 r)/ 2$,
the last of which $\lambda_- = -\half \cos^2 r$,  is
always negative for $0\leq a\leq \infty $ i.e. for $0\leq r\leq \pi
/4$. This means that the state is always entangled.

Quantifying entanglement for mixed states is fairly involved \cite%
{entanglement}. A pure state of a bipartite system $a$ and $b$ can
always be written in the Schmidt basis, $|\psi \rangle
_{ab}=\sum_{n}a_{n}|n\rangle _{a}|n\rangle _{b}$, where the quantum
correlations between the states are evident. Entanglement between
the systems is given by the von-Neumann entropy of the reduced
density matrix $\rho_{a}$ defined as $S(\rho_{a})=-tr(\rho _{a}\log
_{2}(\rho _{a}))$, which is a function of the Schmidt coefficients
$a_{n}$. Unfortunately, there is no analog of the Schmidt
decomposition for mixed states, and the von-Neumann entropy is no
longer a good measure of mixed state entanglement.

A set of conditions that mixed state entanglement measures should
satisfy is well known \cite{entanglement}. There is no unique
measure, and several different mixed state entanglement measures
have been proposed. Among the most popular are those related to the
formation and distillation of entangled states. Consider the number
$m$ of maximally entangled pairs needed to create $n$ arbitrarily
good copies of an arbitrary pure state using only local operations
and classical communication. The entanglement of formation is defined
as the asymptotic conversion ratio, $m/n$
in the limit of infinitely many copies \cite{formation},
\begin{equation*}
E_{F}(\rho _{ab})= \min\sum_{i}p_{i}S(\rho _{a}^{i}),
\end{equation*}%
where the minimum is taken over all the possible realizations of the state $%
\rho _{ab}=\sum_{i}p_{i}|\Psi _{ab}^{i}\rangle \langle \Psi _{ab}^{i}|$.

The opposite process gives rise to the definition of the
entanglement of distillation. This is the asymptotic rate of
converting non-maximally entangled states into maximally entangled
states by means of a purification procedure. The entanglement of
distillation is in general smaller than that of formation. This
shows that the entanglement conversion is irreversible and is due to
a loss of classical information about the decomposition of the
state. Bound entangled states are a consequence of this: no
entanglement can be distilled from them even though they are
inseparable.

The entanglement of formation can be explicitly calculated for two
qubits. It is given by
\begin{eqnarray}
E_{F}=&-&\frac{1+\sqrt{1-C^{2}}}{2}\,\log_{2}\left(\frac{1+\sqrt{1-C^{2}}}{2}\right)  \notag\\
&-&\frac{1-\sqrt{1-C^{2}}}{2}\,\log_{2}\left(\frac{1-\sqrt{1-C^{2}}}{2}\right),
\notag
\end{eqnarray}
where $C$ is the concurrence. The concurrence of a pure state
$\left| \psi \right\rangle $ of two qubits is given by
\begin{equation}
C\left( \psi \right) \equiv \left| \left\langle \psi |\sigma _{y}\otimes
\sigma _{y}|{\psi } ^{*}\right\rangle \right| ,  \notag
\end{equation}
where $\sigma _{y}\otimes \sigma _{y}\left| \psi ^{*}\right\rangle$
represents the `spin-flip' of $\left| \psi \right\rangle $ and ``*"
denotes complex conjugation in the standard basis.

The generalization of the concurrence to a mixed state $\rho $ of two qubits
follows by minimizing the average concurrence over all possible pure state
ensemble decompositions of $\rho $, defined by a convex combination of pure
states $S_{i}=\{p_{i},\psi _{i}\} $, such that $\rho =\sum _{i}p_{i}|\psi
_{i}\rangle \langle \psi _{i}| $. In this way,

\begin{eqnarray}
C\left( \rho \right) &=& \min_{S_{i}}\sum_{i}p_{i}C\left( \psi _{i}\right)
\notag \\
&=&\min_{S_{i}}\sum_{i}p_{i}\left| \left\langle \psi _{i}|\sigma _{y}\otimes
\sigma _{y}|{\psi ^{*} _{i}}\right\rangle \right| .  \notag
\label{Eq Avg Concurrence}
\end{eqnarray}%
%Wootters succeeded in deriving an analytic solution to this
%difficult minimization procedure in terms of the eigenvalues
%$\lambda _{i}$ (all of which are positive) of the non-Hermitian operator $\rho \widetilde{\rho }$,
%where $\tilde{\rho} = (\sigma_y\otimes\sigma_y)\,\rho^*\,(\sigma_y\otimes\sigma_y)$
%is the spin-flip of the quantum state $\rho$ \cite{formation}. The
%closed form solution for the concurrence of a mixed state of two qubits is
%given by
%
Wootters \cite{formation} succeeded in deriving an analytic solution to this
difficult minimization procedure  in terms of the quantities
$\lambda _{i}$ defined as the square roots of the  eigenvalues (which are all positive)
of the non-Hermitian operator $\rho \widetilde{\rho }$,
ordered in decreasing order.
The matrix $\tilde{\rho} = (\sigma_y\otimes\sigma_y)\,\rho^*\,(\sigma_y\otimes\sigma_y)$
is the spin-flip of the quantum state $\rho$  which not only
exchanges the states $|0\rangle$ and $|1\rangle$, but, in general, also introduces a relative phase. The
closed form solution for the concurrence of a mixed state of two qubits is
given by

\begin{equation}  \label{Eq Analytic Mixed Concurrence}
C\left( \rho \right) =\max \left\{ 0,\lambda _{1}-\lambda _{2}-\lambda
_{3}-\lambda _{4}\right\}, \quad \lambda_i\ge \lambda_{i+1}\ge 0. \notag
\end{equation}
Unfortunately, the entanglement of distillation cannot be explicitly
calculated in this case.
Therefore we will use the logarithmic negativity \cite{negativity}
which serves as an upper bound on the entanglement of distillation.
% As we mentioned before, the logarithmic negativity
% is an entanglement monotone that serves as a lower
% bound on the entanglement of distillation.
It is not an entanglement measure because it does not satisfy the requirement
of being equal to the von-Neuman entropy for pure states. However, it is an
entanglement monotone since it satisfies all other criteria to
quantify entanglement. It is defined as
$N(\rho )=\log_{2}||\rho^{T}||_{1}$ where $||\rho ^{T}||_{1}$ is the trace-norm of
the partial transpose density matrix $\rho ^{T}$ i.e., the sum of
the eigenvalues of $\sqrt{(\rho ^{T})^\dagger \rho^T}\,$. Since the
matrices $\rho^T$ in this work are symmetric, $N(\rho )$ is simply given
by $\log_{2}$ of the sum of the absolute values of the eigenvalues of $\rho ^{T}$.

To quantify the entanglement of $\rho_{A,I}$ in Eq.~(\ref{eq:state}) we compute the
spin-flip matrix
$\tilde{\rho}_{A,I}$
\begin{equation*}
\tilde{\rho}_{A,I}=\half \left(
\begin{array}{cccc}
1 & 0 & 0 & \cos r \\
0 & 0 & 0 & 0 \\
0 & 0 & \sin ^{2}r & 0 \\
\cos r & 0 & 0 & \cos ^{2}r%
\end{array}%
\right),  \\
\end{equation*}
and find that
\begin{equation*}
\rho_{A,I}\, \tilde{\rho}_{A,I}=\half \left(
\begin{array}{cccc}
\cos ^{2}r & 0 & 0 & \cos ^{3}r \\
0 & 0 & 0 & 0 \\
0 & 0 & 0 & 0 \\
\cos r & 0 & 0 & \cos ^{2}r%
\end{array}%
\right),
\end{equation*}
has eigenvalues $(\cos ^{2}r,0,0,0)$. The concurrence is then given
by $C(\rho _{A,I})=\lambda _{1}=\cos r$, which
is unity at zero acceleration, as expected, and approaches the value $1/%
\sqrt{2}$ for infinite acceleration $r\rightarrow \pi /4$.
The entanglement of formation is
\begin{eqnarray*}
E_{F}=&-&\frac{1}{2}(1+\sin {r})\log _{2}(\frac{1+\sin {r}}{2}) \no
      &-&\frac{1}{2}(1-\sin {r})\log _{2}(\frac{1-\sin {r}}{2}),
\end{eqnarray*}
and the logarithmic negativity is
$$N=\log_{2}(1+\cos ^{2}r)
$$

In an inertial frame, $r=0$, and the state of the system defined by
\Eq{eq:ent} is maximally entangled. In
the limit of infinite acceleration, the logarithmic negativity is $\log _{2}{%
\frac{3}{2}}=0.585$, implying that the entanglement in the infinite
acceleration limit is finite. This means that the state is always
entangled and can be used as a resource for performing certain
quantum information processing tasks. This is in strong contrast to
the bosonic case \cite{ivette}, where it is found that the
entanglement goes to zero in this limit. As in \cite{ivette} the
infinite acceleration limit can be interpreted as Alice falling into
a black hole while Rob hovers outside the black hole, barely
escaping the fall. Since close to a black hole horizon spacetime is
flat, Rob must be uniformly accelerated in order to escape the black
hole. However, Alice falls in as she is an inertial observer.  The
connection to the present situation is made by recognizing that,
according to Rob, a communication horizon appears causing him to
lose information about the state in the whole of spacetime. As a
result, the inertial entanglement is degraded, and there will be a
reduction in the fidelity of any information processing task
performed by Alice and Rob using this state. The analogy with the
black hole scenario is further strengthened by observing that
classical information can only flow from Rob to Alice after Alice
has crossed the horizon.

We may also calculate the total correlations between
any two subsytems of the overall system by using
the mutual information
\cite{mutual},
\begin{equation*}
I=S(\rho _{a})+S(\rho _{a})-S(\rho _{ab}).  \notag
\end{equation*}%
This measure quantifies how much information two correlated
observers (one with access to subsystem $a$ and the other with
access to subsystem $b$) possess about one another's state.
Equivalently, it represents the distance between the actual joint
distribution and the product state obtained when all correlations
are neglected. Calculating the relevant marginal density operators,
the mutual information of the state (\ref{eq:state}) is found to be
\begin{eqnarray}
I &=&1-\frac{1}{2}\cos ^{2}r  \,\log _{2}\left(\frac{\cos ^{2}r}{2}\right) \\
  &-&(1-\frac{1}{2}\cos ^{2}r)\,\log _{2}\left(1-\frac{1}{2}\cos ^{2}r\right)\notag \\
  &+&\frac{1}{2}(1+\cos ^{2}r)\,\log _{2}\left(\frac{1+\cos ^{2}r}{2}\right)\notag \\
  &+&\frac{1}{2}(1-\cos ^{2}r)\,\log _{2}\left(\frac{1-\cos ^{2}r}{2}\right),\notag
\end{eqnarray}%
where we have used the fact that the density matrix in region I is
$\rho_{I}=\frac{1}{2}\left[\cos ^{2}r|0\rangle_{I}\langle 0|
+(\sin^{2}r+1)|1\rangle_{I}\langle 1|\right]$,
so that
\begin{eqnarray}  \label{sr1}
S(\rho_{I})= &-&\frac{1}{2}\cos ^{2}r   \,\log_{2}\left(\frac{\cos^{2}r}{2}\right) \\
             &-&(1-\frac{\cos ^{2}r}{2})\,\log_{2}\left(1-\frac{\cos ^{2}r}{2}\right).
\notag
\end{eqnarray}%
Similarly, the density matrix of mode $A$ is
$\rho_{A}=\frac{1}{2}(|0_A\rangle\langle0_A| +|1_A\rangle\langle 1_A|)$,
yielding $S(\rho_{A})=1$.  Finally, using \Eq{eq:state} we
find that
\begin{eqnarray}
S(\rho _{A,I})= &-&\frac{1}{2}(1+\cos ^{2}r)\,\log_{2}\left(\frac{1+\cos^{2}r}{2}\right)  \label{skr1} \\
                &-&\frac{1}{2}(1-\cos ^{2}r)\,\log_{2}\left(\frac{1-\cos ^{2}r}{2}\right). \notag
\end{eqnarray}
The mutual information is equal to $2$ at $r=0$ and goes to unity in
the infinite acceleration limit. This behavior is reminiscent of
that seen in the bosonic case \cite{ivette}.

\section{Entanglement in other partitions and Entanglement sharing}\label{sec:ent_sharing}

To explore entanglement in this system in more detail we consider
the tripartite system consisting of the modes $A$, I, and II. In an
inertial frame the system is bipartite, but from a non-inertial
perspective an extra set of modes in region II becomes relevant. We
therefore calculate the entanglement in all possible bipartite
divisions of the system as well as any tripartite correlations that
may exist.

\subsection{Pure state entanglement}

We investigate pure state entanglement in three different bipartite
divisions of the system. Since the overall state of the system is
pure, these entanglements are uniquely quantified by the von-Neumann
entropy.  In what follows we make repeated use of the fact
that, for a bipartite pure state $\rho_{ab}$ with subsystems $a$
and $b$, $S(\rho_{ab})=0$ and $S(\rho _{a})=S(\rho _{b})$.
There are three separate cases to consider.

\begin{center}
\begin{figure}[tbp]
\label{entropy} \epsfig{file=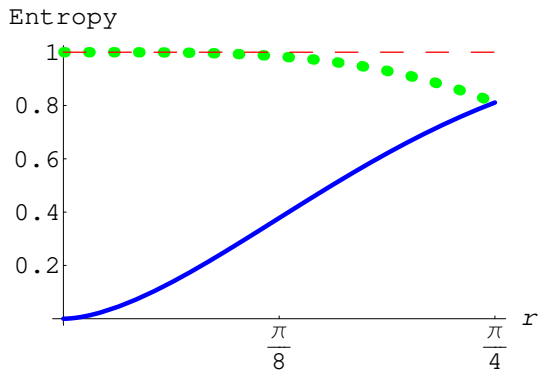, width=8cm}
\caption{(Color online) Bipartite pure state entanglement. Thick
solid curve (Blue): Anti-Rob in region II with (Alice and Rob in
region I). Thin dashed curve (Red): Alice with (regions I and II).
Dotted curve (Green): Rob in region I with (Alice and anti-Rob
region II.) } \label{fig:entropy}
\end{figure}
\end{center}

\begin{itemize}
\item The entanglement between mode $A$ and (modes I and II) is
given by $S(\rho_{A})=S(\rho_{I,II})=1$. Thus, these two subsystems
are always maximally entangled, regardless of the value of the
parameter $r$.

\item The entanglement between mode I and (modes $A$ and II)
is given by Eq.~(\ref{sr1}) since $S(\rho_{I})=S(\rho _{A,II})$.

\item Finally, the entanglement between mode II and (modes $A$ and I)
is given by Eq.~(\ref{skr1}) since $S(\rho_{II})=S(\rho _{A,I})$.
\end{itemize}
Figure~(\ref{fig:entropy}) shows the entanglement in each of these
bipartite partitions as functions of $r$.

\subsection{Mixed state entanglement}

For mixed state entanglement we can consider, besides the
entanglement between the Minkowski mode $A$ and Rindler mode I
calculated in Sec.~\ref{sec:ent}, two additional bipartite divisions
of the subsystems.

\begin{figure}[tbp]
\label{nega} \epsfig{file=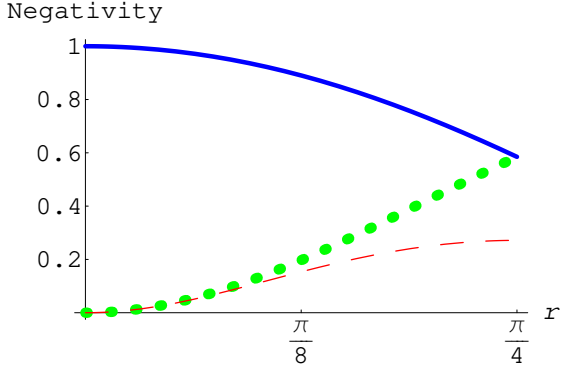,width=8cm} \caption{(Color
online) The logarithmic negativity as a function of $r$. Thick solid
curve (Blue): between Alice and Rob in region I. Thin dashed curve
(Red): between the modes in regions I and II. Dotted curve (Green):
between Alice and anti-Rob in region II.} \label{fig:nega}
\end{figure}

\begin{itemize}
\item The entanglement between mode $A$ and mode II.
Tracing over the mode in region I, we obtain the density matrix
\begin{equation*}
\rho_{A,II}=\frac{1}{2}\left(
\begin{array}{cccc}
\cos ^{2}r & 0 & 0 & 0 \\
0 & \sin ^{2}r & \sin {r} & 0 \\
0 & \sin {r} & 1 & 0 \\
0 & 0 & 0 & 0%
\end{array}%
\right).
\end{equation*}%
The partial transpose of $\rho_{A,II}$ is given by
\begin{equation*}
\rho^T_{A,II}=\frac{1}{2}\left(
\begin{array}{cccc}
\cos ^{2}r & 0 & 0 & \sin {r} \\
0 & \sin ^{2}r & 0 & 0 \\
0 & 0 & 1 & 0 \\
\sin {r} & 0 & 0 & 0%
\end{array}%
\right),
\end{equation*}%
which has eigenvalues $(1,1,\sin^2 r, -\sin^2 r)/ 2$,
the last of which $\lambda_- = -\half \sin^2 r$, is less than or equal to zero.
At $r =0$ the eigenvalue is zero, which means that
there is no entanglement at this point. However, for any $r>0$
entanglement does exist between these two modes
% since the negativity in this case is given by $N=\log _{2}(1+\sin^{2} r)$.
according to the partial transpose criterion. The logarithmic negativity
in this case is given by $N=\log _{2}(1+\sin^{2} r)$.

Calculating the spin-flip of $\rho_{A,II}$
\begin{equation}
\tilde{\rho}_{A,II}=\frac{1}{2}\left(
\begin{array}{cccc}
0 & 0 & 0 & 0 \\
0 & 1 & \sin{r} & 0 \\
0 & \sin {r} & \sin^{2}{r} & 0 \\
0 & 0 & 0 & \cos^{2}{r}%
\end{array}%
\right),  \notag
\end{equation}%
we find that
\begin{equation}
\rho_{A,II}\,\tilde{\rho}_{A,II}=\frac{1}{2}
\left(
\begin{array}{cccc}
0 & 0 & 0 & 0 \\
0 & \sin ^{2}{r} & \sin ^{3}{r} & 0 \\
0 & \sin {r} & \sin ^{2}{r} & 0 \\
0 & 0 & 0 & 0%
\end{array}%
\right),  \notag
\end{equation}%
has eigenvalues $(\sin^{2}r,0,0,0)$.
Thus, the concurrence is given by
$C(\rho _{A,II})=\lambda _{1}=\sin r $, which is zero at zero acceleration as
expected, and approaches the value $1/\sqrt{2}$ for infinite acceleration
$r\rightarrow \pi /4$. The entanglement of formation is
\begin{eqnarray*}
E_{F}= &-&\frac{1}{2}(1+\cos {r})\log _{2}\left(\frac{1+\cos {r}}{2}\right) \\
&-&\half(1-\cos {r})\log _{2}\left(\frac{1-\cos {r}}{2}\right)
\end{eqnarray*}%
and the mutual information is%
\begin{eqnarray}
I &=&S(\rho _{A})+S(\rho _{II})-S(\rho _{A,II})  \notag \\
  &=&1-\frac{1}{2}(1+\cos ^{2}r)\log _{2}\left(\frac{1+\cos ^{2}r}{2}\right)\notag \\
&-&\frac{1}{2}(1-\cos ^{2}r)\log _{2}\left(\frac{1-\cos ^{2}r}{2}\right)\notag \\
&+&\frac{1}{2}\cos ^{2}r\log _{2}\left(\frac{\cos ^{2}r}{2}\right)  \notag \\
&+&(1-\frac{1}{2}\cos ^{2}r)\log _{2}\left(1-\frac{1}{2}\cos^{2} r\right). \notag
\end{eqnarray}%
At $r=0$ the mutual information is zero and approaches unity as the
rate of acceleration goes to infinity.

\item The entanglement between mode I and mode II. Tracing over the modes
in $A$, we obtain the density matrix
\begin{equation*}
\rho _{I,II}=\frac{1}{2}\left(
\begin{array}{cccc}
\cos ^{2}r & 0 & 0 & \cos {r}\sin {r} \\
0 & 0 & 0 & 0 \\
0 & 0 & 1 & 0 \\
\cos {r}\sin {r} & 0 & 0 & \sin ^{2}r%
\end{array}%
\right).
\end{equation*}%
The partial transpose of $\rho _{I,II}$ (obtained by interchanging $I$'s qubits)
is given by
\begin{equation*}
\rho^T_{I,II}=\frac{1}{2}\left(
\begin{array}{cccc}
\cos ^{2}r & 0 & 0 & 0 \\
0 & 0 & \cos {r}\sin {r} & 0 \\
0 & \cos {r}\sin {r} & 1 & 0 \\
0 & 0 & 0 & \sin ^{2}r%
\end{array}%
\right).
\end{equation*}%
which has eigenvalues $(2 \sin^2 r, 2 \cos^2 r,1 + (1+\sin^2 2r)^{1/2}, 1-(1+\sin^2 2r)^{1/2})/ 4$,
the last of which $\lambda_- = -\frac{1}{4}[1-\sqrt{1+\sin^2 2r}]$
is less than or equal to zero.
Again, at $r=0$ there is no
entanglement between these two subsystems.  Yet, similar to the last
case, entanglement does exist between these two modes in non-inertial frames
according to the partial transpose criterion.
%since the negativity $N=\log _{2}(\frac{1}{2}({1+\sqrt{1+\sin^{2} 2r}}))$ is nonzero for all $r>0$.
Further, the logarithmic negativity
$N=\log _{2}(\frac{1}{2}({1+\sqrt{1+\sin^{2} 2r}}))$ is nonzero for all $r>0$.
The spin-flip of $\rho_{I,II}$
is given by
\begin{equation}
\tilde{\rho}_{I,II}=\frac{1}{2}
\left(
\begin{array}{cccc}
\sin ^{2}r & 0 & 0 & \cos r\sin {r}\\
0 & 1 & 0 & 0 \\
0 & 0 & 0 & 0 \\
\cos {r}\sin r & 0 & 0 & \cos ^{2}r %
\end{array}%
\right),  \notag
\end{equation}%
and the matrix
\begin{equation}
\rho_{I,II}\,\tilde{\rho}_{I,II}=\frac{1}{2}
\left(
\begin{array}{cccc}
\cos ^{2}r\sin ^{2}r & 0 & 0 & \cos ^{3}r\sin {r}\\
0 & 0 & 0 & 0 \\
0 & 0 & 0 & 0 \\
\cos {r}\sin ^{3}r & 0 & 0 & \cos ^{2}r\sin ^{2}r%
\end{array}%
\right),  \notag
\end{equation}%
has eigenvalues
$(\cos ^{2}r\sin^{2}r,0,0,0)$. \ Thus, the concurrence is given by
$C(\rho _{I,II})=\lambda _{1}=\sin {r}\cos {r}$ which is zero at zero
acceleration, and approaches the value $1/{2}$ for infinite
acceleration $r\rightarrow \pi /4$. The entanglement of formation in
this case is
\begin{widetext}
\begin{eqnarray}
E_{F} =&-&\frac{(1+\sqrt{1-\sin^2{r}\cos^2{r}})}{2}
\,\log _{2}\left(\frac{1+\sqrt{1-\sin^2{r}\cos^2{r}}}{2}\right) \\
&-&\frac{(1-\sqrt{1-\sin^2{r}\cos^2{r}})}{2}\,\log _{2}\left(\frac{1-\sqrt{1-\sin^2{r}\cos^2{r}}}{2}\right),  \notag
\end{eqnarray}
\end{widetext}
and the mutual information is%
\begin{eqnarray}
I &=&S(\rho _{I})+S(\rho _{II})-S(\rho _{I,II})  \notag \\
&=&-\frac{1}{2}\cos ^{2}r    \,\log _{2}\left(\frac{\cos ^{2}r}{2}\right)\notag \\
&-&(1-\frac{1}{2}\cos ^{2}r) \,\log _{2}\left(1-\frac{1}{2}\cos ^{2}r\right)\notag \\
&-&\frac{1}{2}(1+\cos ^{2}r) \,\log _{2}\left(\frac{1+\cos ^{2}r}{2}\right)\notag \\
&-&\frac{1}{2}(1-\cos ^{2}r) \,\log _{2}\left(\frac{1-\cos ^{2}r}{2}\right)-1.\notag
\end{eqnarray}
\end{itemize}
Again, we find that the mutual information is zero at $r=0$, and
increases to a finite value (in this case
$I=3/2\left(2-\log_{2}3\right)$), as the rate of acceleration goes
to infinity.  The results of the above calculations are shown in
Figs.~(\ref{fig:nega}) - (\ref{fig:mutual}).

\begin{center}
\begin{figure}[tbp]
\label{formation} \epsfig{file=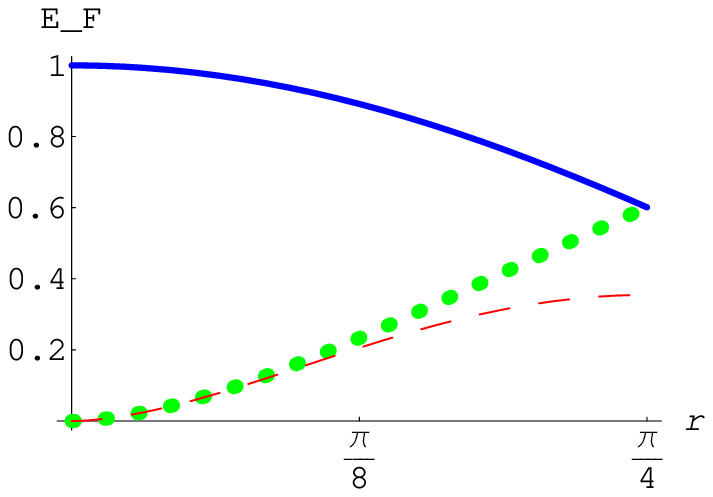,width=8cm}
\caption{(Color online) Entanglement of formation as a function of
$r$. Thick solid curve (Blue): between Alice and Rob in region I.
Thin dashed curve (Red): between the modes in regions I and II.
Dotted curve (Green): between Alice and anti-Rob in region II.}
\label{fig:formation}
\end{figure}

\begin{figure}[tbp]
\label{mutual} \epsfig{file=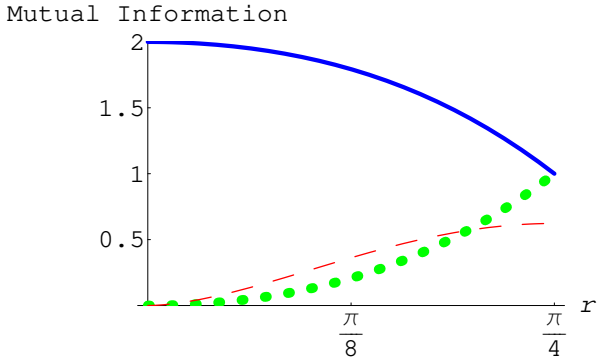, width=8cm} \caption{(Color
online) Mutual information as a function of $r$. Thick solid curve
(Blue): between Alice and Rob in region I. Thin dashed curve (Red):
between the modes in regions I and II. Dotted curve (Green): between
Alice and anti-Rob in region II.} \label{fig:mutual}
\end{figure}
\end{center}

\subsection{Tripartite entanglement}

Entanglement in triparite systems has been studied by Coffman, et.
al., \cite{Coffman} for the case of three qubits. They found that
such quantum correlations cannot be arbitrarily distributed amongst
the subsystems; the existence of three-body correlations constrains
the distribution of the bipartite entanglement which remains after
tracing over any one of the qubits. For example, in a GHZ-state,
$|\mathrm{GHZ}\rangle =|000\rangle +|111\rangle $, tracing over one
qubit results in maximally mixed state containing no entanglement
between the remaining two qubits. In constrast, for a W-state,
$|\mathrm{W}\rangle =|001\rangle +|010\rangle +|100\rangle $, the
average remaining bipartite entanglement is maximal. Coffman, et.
al., analyzed this phenomenon of entanglement sharing \cite{Coffman},
using an entanglement monotone known as the tangle, defined as the
square of the concurrence $\tau =C^{2}$ . They also introduced a new
quantity, known as the residual tangle, in order to quantify the
irreducible tripartite correlations in a system of three qubits
($a$, $b$, and $c$) \cite{Coffman}. The definition is motivated by
the observation that the tangle of $a$ with $b$ plus the tangle of
$a$ with $c$ cannot exceed the tangle of $a$ with the joint
subsystem $bc$, i.e.,

\begin{equation}  \label{Eq Tangle Ineq}
\tau _{a,b}+\tau _{a,c}\leq \tau _{a\left( b,c\right) }.
\end{equation}

Subtracting the terms on the left hand side of Eq. (\ref{Eq Tangle
Ineq}) from that on the right hand side yields a nonnegative
quantity referred to as the residual tangle $\tau _{a,b,c} $, i.e.,

\begin{equation}  \label{Eq Residual Tangle}
\tau _{a,b,c}\equiv \tau _{a\left( b,c\right) }-\tau _{a,b}-\tau
_{a,c}.
\end{equation}
The residual tangle or 'three tangle' is interpreted as quantifying the inherent
tripartite entanglement present in a system of three qubits, i.e.,
the entanglement that cannot be accounted for in terms of the
various bipartite tangles. This interpretation is given further
support by the observation that the residual tangle is invariant
under all possible permutations of the subsystem labels
\cite{Coffman}.

To quantify tripartite entanglement in our system we use the
residual tangle $\tau _{A,I,II}=\tau _{A(I,II)}-\tau _{A,I}-\tau
_{A,II}$. This quantity is zero for the situation we are
considering,
since $\tau _{A,I}=\cos ^{2 }r$, $%
\tau _{A,II}=\sin ^{2 }r$ and $\tau _{A(I,II)}=2(1-Tr[\rho
_{I,II}^{2}])=1$. Thus, the state that we are studying has no
tripartite correlations for any value of the acceleration rate.
Instead, any entanglement existing in the system is necessarily
bipartite in nature.

The marginal bipartite tangles, plotted in Fig.~\ref{fig:tangles} as
functions of $r$, are found to be strongly constrained. For low
rates of acceleration, modes $A$ and I remain almost maximally
entangled while there is very little entanglement between modes I
and II and between modes $A$ and II. As the acceleration grows, the
entanglement between modes I and II and between modes $A$ and II
increases, while the entanglement between modes $A$ and I is
degraded. The main system of interest (mode $A$ plus mode I) becomes
increasingly entangled to mode II and therefore, after tracing over
mode II, we observe an effect analogous to environmental decoherence
in which the modes in region II play the role of the environment.
The distribution of entanglement in the system is non-trivial since
entanglement between modes $A$ and I is not conserved as it is in
the inertial case.

\section{Complementarity}\label{sec:complementarity}

We next analyze our system in terms of several complementarity
relations designed to identify the different types of information
encoded in a quantum state in an attempt to better understand how
various subsystem properties depend on Rob's rate of acceleration.
Specifically, we are interested in explaining (i) the
nonconservative nature of the entanglement discussed in the last
section and (ii) the fact that not all of the initial entanglement
between Alice and Rob is destroyed, even at infinite acceleration.
The latter of these two results corresponds to the most obvious
difference between the fermionic case studied here and the bosonic
case investigated in \cite{ivette}.

We begin by making use of the relationship \cite{Tessier}
\begin{equation}
\eta\left(\rho \right)+\tau \left(\rho
\right)+\overline{S^{2}}\left(\rho_{a}\right)
+\overline{S^{2}}\left(\rho_{b}\right)=1,\label{eq:complementarity}
\end{equation}
which shows that an arbitrary state of two qubits ($a$ and $b$)
exhibits a complementary tradeoff between the amounts of separable
uncertainty $\eta$, bipartite entanglement (as quantified by the
tangle) $\tau$, and a unitarily invariant measure of information
about the single particle properties $\overline{S^2}$ that it
encodes. The separable uncertainty \begin{equation} \eta \left(\rho
\right)\equiv \mbox{\rm Tr}\left(\rho \tilde{\rho}
\right)+M\left(\rho \right)-\tau \left(\rho
\right),\label{eq:LocalCorrelations}\end{equation} $0\leq
\eta\left(\rho\right) \leq 1,$ is a measure of the uncertainty or
ignorance encoded in the two-qubit mixed state $\rho$ regarding
individual subsystem properties that is unrelated to the presence of
entanglement between the qubits.  In
Eq.~(\ref{eq:LocalCorrelations}), $\tilde{\rho}$ represents the
spin-flip of $\rho$, and $M\left(\rho \right)\equiv 1-\mbox{\rm
Tr}\left(\rho^{2}\right)$ is the marginal mixedness of $\rho$.

Similarly, $\overline{S^2}\left(\rho_k\right)$ is a measure of the
information pertaining to a single qubit encoded in the marginal
density operator $\rho_k$.  Specifically,
$\overline{S^{2}}\left(\rho_{k} \right)\equiv
1/2\left[\nu^{2}\left(\rho_{k}\right)+p^{2}\left(\rho_{k}\right)\right]$
is the average of the squares of the single qubit properties
associated with qubit $k=a,b$. The first of these properties, the
coherence $\nu$ of qubit $k$, quantifies, e.g., the fringe
visibility in the context of a two-state system incident on an
interferometer, and is given by
\begin{equation}
\nu \left(\rho_{k}\right)\equiv 2\left|\mbox{\rm Tr}\left(\rho
_{k}\sigma
_{+}^{\left(k\right)}\right)\right|,\label{eq:Coherence}\end{equation}
where $\sigma _{+}^{\left(k\right)}=\left(\begin{array}{cc}
 0 & 1\\
 0 & 0\end{array}
\right)$ is the raising operator acting on qubit $k$. Similarly, the
predictability $p$ which quantifies the \textit{a priori}
information regarding whether qubit $k$ is in the state
$\left|0\right\rangle $ or the state $\left|1\right\rangle $, e.g.,
whether it is more likely to take the upper or lower path in an
interferometer, is given by\begin{equation}
p\left(\rho_{k}\right)\equiv \left|\mbox{\rm Tr}\left(\rho
_{k}\sigma
_{z}^{\left(k\right)}\right)\right|,\label{eq:Predictability}\end{equation}
where $\sigma _{z}=\left(\begin{array}{cc}
 1 & 0\\
 0 & -1\end{array}
\right)$ and $\left|0\right\rangle \left(\left|1\right\rangle
\right)$ is the plus (minus) one eigenvector of $\sigma _{z}$.

\begin{figure}[h]
\label{tangles}
\begin{center}
\epsfig{file=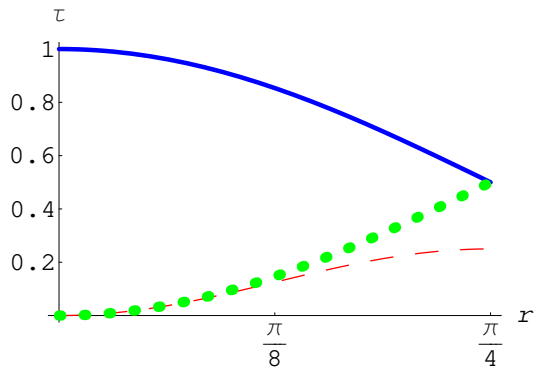, width=8cm}
\end{center}
\caption{(Color online) Bipartite tangles as a function of $r$.
Thick solid curve (Blue): between Alice and Rob in region I. Thin
dashed curve (Red): between the modes in regions I and II. Dotted
curve (Green): between Alice and anti-Rob in region II.}
\label{fig:tangles}
\end{figure}

\begin{figure}[h]
\label{eta}
\begin{center}
\epsfig{file=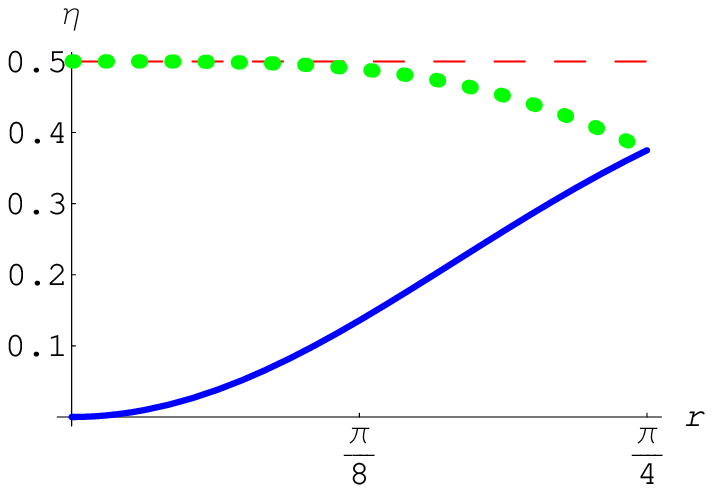, width=8cm}
\end{center}
\caption{(Color online) Separable uncertainties as a function of
$r$. Thick solid curve (Blue): between Alice and Rob in region I.
Thin dashed curve (Red): between the modes in regions I and II.
Dotted curve (Green): between Alice and anti-Rob in region II.}
\label{fig:eta}
\end{figure}

\begin{figure}[h]
\label{single_part}
\begin{center}
\epsfig{file=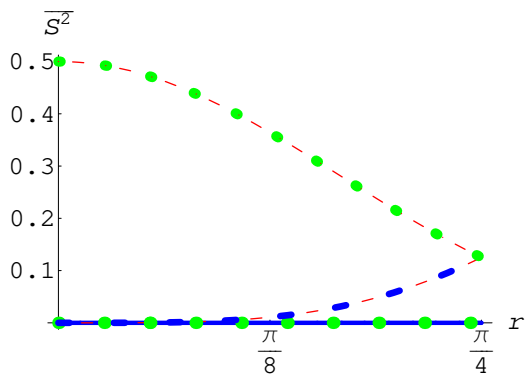, width=8cm}
\end{center}
\caption{(Color online) Single qubit properties as a function of
$r$. Dotted-solid line (Blue-Green): Alice (always zero). Thick-thin
dashed curve (Blue-Red): Rob in region I. Dot-dashed curve
(Green-Red): anti-Rob in region II.} \label{fig:single_part}
\end{figure}

Figures~\ref{fig:tangles} and \ref{fig:eta} plot the bipartite
components of Eq.~(\ref{eq:complementarity}) for the various
two-qubit marginals obtained after tracing over any one of the three
subsystems ($A$, I, or II) as functions of the parameter $r$, while
Fig.~\ref{fig:single_part} does the same for the relevant single
particle information measures. The complementary nature of these
quantities is illustrated by the pattern/color scheme chosen for the
curves, where each pattern (color) corresponds to a unique set of
properties satisfying Eq.~(\ref{eq:complementarity}). Adding
together all of the curves that contain a common attribute in
Figs.~\ref{fig:tangles} - \ref{fig:single_part} yields the constant
value one, independent of $r$.
For example, taking the sum of the two dashed (red) curves in
Figs.~\ref{fig:tangles} and \ref{fig:eta} and the two curves
containing dashes (red) in Fig.~\ref{fig:single_part}, thick-thin
dashed (blue-red) and dot-dashed (green-red), corresponds to the
equality $\tau_{I,II}+\eta_{I,II}
+\overline{S^2}\left(\rho_{I}\right)
+\overline{S^2}\left(\rho_{II}\right)=1$. Note that the curves in
Fig.~\ref{fig:single_part}, representing information encoded in the
individual qubits, each contribute to exactly two distinct
complementarity relations.

The tradeoffs expressed by Eq.~(\ref{eq:complementarity}), and
illustrated in Figs.~\ref{fig:tangles} - \ref{fig:single_part}, show
how Rob's acceleration determines the distribution of individual and
bipartite properties of the system. For example, adding together the
thick solid (blue) and dotted (green) curves in
Fig.~\ref{fig:tangles} shows that, although the entanglement in any
bipartite subsystem changes with acceleration, the total entanglement
between Alice and (the modes in regions I and II) is always maximal
and constant.  Further, since Rob is forced to trace over the
causally disconnected modes in region II, the entanglement arising
between the modes in region I and region II is ultimately
responsible for the Unruh radiation that he sees. Indeed Alice, who
has access to the modes in both regions, always sees an undisturbed
vacuum. However, our analysis shows that the maximum amount of
thermalization that Rob can experience (even at infinite
acceleration) is limited by the amount of entanglement that exists
between Alice and Rob in an inertial frame.

One straightforward way to see this is to consider the case in which
Alice and Rob initially share no entanglement (e.g. an arbitrary product state between
Alice and Rob). Then, as Rob's acceleration approaches infinity, we find that $\tau_{I,II}$
approaches its maximum possible value of one (see \Eq{vacM} with $r=\pi/4$ \cite{note4}).
This is in contrast to the situation considered here in which the initial entanglement
between Alice and Rob is maximal, constraining $\tau_{I,II}$ to a
maximum value of 1/4.

Since the overall state of our system is always pure, we may also
apply the result \cite{Peng}
\begin{equation}
\tau_{a,b,c}+\tau^{\left(k\right)}+\overline{S^{2}}\left(\rho_{k}\right)=1,
\label{eq:EntSharingRel}
\end{equation}
which holds for an arbitrary pure state of three qubits ($a$, $b$,
and $c$). The quantity $\tau^{\left(k\right)}\equiv \sum_{j\neq
k}\tau_{j,k}$ is a measure of the total pairwise entanglement shared
between qubit $k$ and all other qubits.
Equation~(\ref{eq:EntSharingRel}) quantifies an explicit tradeoff
between the inherent three-body correlations in a tripartite system,
the total pairwise entanglement of a selected qubit, and the single
particle properties of the qubit.

Because the three tangle \Eq{Eq Residual Tangle} is always zero for our system, this pure
state complementarity relationship simplifies to
\begin{equation}
\tau^{\left(k\right)}+\overline{S^{2}}\left(\rho_{k}\right)=1.
\label{eq:EntSharingRelSimple}
\end{equation}
It is then straightforward to see that this tradeoff is also
captured by the pattern/color scheme used in the above figures.
Simply adding the sum of any two curves in Fig.~\ref{fig:tangles} to
the curve in Fig.~\ref{fig:single_part} that is composed of the same
two attributes (or colors) again yields the constant value one,
regardless of Rob's acceleration. In particular, this shows that the
entanglement between Alice and Rob in region I cannot vanish at
infinite acceleration as it did in the bosonic case.  This is
because the entanglement $\tau_{I,II}$ generated between regions I and II
(dashed/red curve in Fig.~\ref{fig:tangles}) and the single particle
properties $\overline{S^{2}}\left(\rho_{I}\right)$
that manifest for Rob (dashed/red-blue curve in
Fig.~\ref{fig:single_part}) are insufficient to satisfy
Eq.~(\ref{eq:EntSharingRelSimple}). Instead, the contribution from
the entanglement $\tau_{A,I}$ between Alice and Rob (solid/blue curve in
Fig.~\ref{fig:tangles}) must also be taken into account to ensure
that
$\tau_{A,I}+\tau_{I,II}+\overline{S^{2}}\left(\rho_{I}\right)=1$.

Finally, we note that the various two-qubit marginal density
operators for our system have a very specific form; they are all
examples of the maximally entangled states with fixed marginal
mixednesses (MEMMS) identified in \cite{Adesso}.  Indeed, this fact
is closely related to the absence of tripartite correlations in our
system, which implies that any entanglement that exists in the
system must necessarily be bipartite.  As shown in \cite{Tessier},
the MEMMS are characterized by the relationship
$M\left(\rho\right)=\eta\left(\rho\right)$, i.e., the marginal
mixednesses are equal to the separable uncertainties.  Thus, our
fermionic system encodes the maximum amount of bipartite
entanglement consistent with the separable uncertainty in each
two-qubit marginal for every possible value of the acceleration
rate.

\section{Summary and Future Directions}\label{sec:conclusions}

We have studied the behavior of the entanglement between two modes
of a free Dirac field in a non-inertial frame in flat spacetime from
the point of view of two observers, Alice and Rob, in relative
uniform acceleration.  Our results show that entanglement existing
between Alice and Rob in an inertial frame is progressively degraded
by the Unruh effect as Rob's rate of acceleration increases.
However, unlike the bosonic case in which the inertial entanglement
vanishes in the limit of infinite acceleration, in this case the
entanglement achieves a minimum value of $1/\sqrt{2}$ for the
entanglement of formation (and of $1/2$ for the tangle). This
fundamental difference, a consequence of the fact that fermions have
access to only two quantum levels vs. the infinite ladder of
excitations available to bosons, means that in this case Alice and
Rob always share some entanglement which can in principle be used as
a resource for performing certain quantum information processing
tasks.  Further analysis shows that the total (quantum plus
classical) correlations in the system, as quantified by the mutual
information, behaves in a manner reminiscent of the bosonic case,
decreasing from 2 for inertial observers to unity in the case of
infinite acceleration.

Considering the (causally inaccessible to Rob) modes in region II to
be a third subsystem allows us to analyze this system in terms of
entanglement sharing. In doing so, we find that the overall
tripartite pure state never encodes any inherently three-body
correlations, regardless of the rate of acceleration. Any
entanglement existing in the fermionic system is therefore
necessarily bipartite. Such entanglement is known to be an invariant
quantity for inertial observers, although different inertial
observers may see these quantum correlations distributed among
multiple degrees of freedom in different ways. However, in this
analysis we find no indication that the entanglement in any
bipartite subsystem is conserved when Rob is allowed to accelerate.
In fact, our results show that the presence of the communication
horizon, which isolates the modes in region II from an accelerated
observer in region I, plays a key role in degrading the inertial entanglement
between Alice and Rob.

Further insight into this behavior is gained by applying both pure
and mixed state complementarity relations to different divisions of
the system into subsystems.  The ability of this formalism to
identify the different types of information encoded in a quantum
state facilitates the study of how various subsystem properties
depend on Rob's rate of acceleration. For instance, the constraints
imposed by these relations illustrate how multi-qubit
complementarity prevents the entanglement between Alice and Rob from
vanishing at infinite acceleration as it does in the bosonic case.
Additionally, we find that the amount of vacuum thermalization (due
to Unruh radiation) that Rob experiences at infinite acceleration is
constrained by the amount of entanglement that he shares with Alice
in an inertial frame.

One possible avenue for further research along these lines is to
study the entanglement between Alice and Rob
in the case that Alice has acceleration $a_{1}$ and Rob has acceleration $%
a_{2}$. In this case the density matrix, after tracing over region
II is
\begin{widetext}
\begin{equation*}
\rho _{A,I}=\frac{1}{2}\left(
\begin{array}{cccc}
\cos ^{2}{r_{1}\cos ^{2}{r_{2}}} & 0 & 0 & \cos {r_{1}}%
\cos {r_{2}} \\
0 & \cos ^{2}{r_{1}}\sin ^{2}{r_{2}} & 0 & 0 \\
0 & 0 & \sin ^{2}{r_{1}}\cos ^{2}{r_{2}} & 0 \\
\cos {r_{1}}\cos {r_{2}} & 0 & 0 & 1+\sin ^{2}{r_{1}}%
\sin ^{2}{r_{2}}%
\end{array}%
\right)
\end{equation*}
\end{widetext}
where $\tan {r_{1}}=\exp (-\omega c/a_1)$ and $\tan {r_{2}}=\exp \left( -\omega c/a_{2}\right)$.
In this case the
logarithmic negativity is $N=\log _{2}(1+\cos ^{2}{r_{1}}\cos ^{2}{%
r_{2}})$.  Thus, we see that the entanglement is further degraded by
having two accelerated observers, but again remains finite, in this
case taking on the value $N=\log _{2}(5/4)$ in the infinite
acceleration limit.

Another aspect of this problem that deserves further consideration
is the nature of the communication that is and is not allowed
between observers in different regions.  Of course, no communication
is possible between the left and right Rindler wedges (regions I and
II). However, even when two parties are not causally disconnected
from one another (e.g. Alice and Rob in \Fig{fig1}), the presence of
a horizon due to the acceleration of an observer imposes a
unidirectionality on any classical communication occurring after the
inertial observer crosses the horizon.  Given the connection between
the horizons seen by accelerated observers, and the event horizon of
a black hole, this in turn suggests the potential for gaining
further insight into questions related to the black hole information
paradox.

%%%%%%%%%%%%%%%%%%%%%%%%%%%%%%%%%%%%%%%%%%%%%%%%%%%%%%%%%%%%%%%%%%%%%%%%%%%%%%%%%%%%%%%%%%%%%%%%%%%%

%\noindent
\acknowledgements
%\medskip
This work was supported in part by the Natural Sciences \& Engineering Research
Council of Canada.


\begin{thebibliography}{99}
\bibitem{note1} To cover the Minkowski regions III and IV in \Fig{fig1} one needs to switch
$\cosh \leftrightarrow \sinh$ in \Eq{Rindler_coords}, which changes the
Rindler metric in these regions to the form
$ds^2 =$ $dt^2 - dz^2 - d^2\mathbf{x}_{\perp} =$
$\exp(2a\zeta)(d^2\zeta - d^2\tau)  - d^2\mathbf{x}_{\perp}$.
This makes $\tau$ a spacelike coordinate and $\zeta$ a timelike coordinate in region III and IV.
We will not require the solutions to the Dirac equations in regions III and IV
in order to discuss the entanglment between Alice and observers in region I and II.
\bibitem{note_on_spin} In this work, we are implicitly assuming that the spin of each mode
is along the direction of acceleration, the $z$-axis.
Since this motion is rectilinear, the succesive boosts into Rob's instantaneous co-moving frame all commute.
Hence, the spin does not suffer any Thomas precession (Wigner rotation) and
therefore points along the same direction in both the inertial and non-inertial frames. See
also \cite{rocio}.
\bibitem{note2} In an arbitrary metric $g_{\mu\nu}(x)$ the Dirac inner product is given by
$\big(\phi(x),\varphi(x) \big)_{\Sigma} = \int d\Sigma_{\mu} \, \bar{\phi}(x) \gamma^{\mu}(x) \varphi(x)$
where where $d\Sigma_{\mu} = n^{\mu} dS$ is a spacelike hypersurface with normal $n^{\mu}$
and hyperspatial volume element $dS$,
and $\bar{\phi}(x) = \phi^{\dagger}(x) \gamma_{M}^{0}$ is the adjoint of the Dirac spinor
wave function $\bar{\phi}(x)$. Here $\gamma_{M}^{\mu}$ are the flat spacetime Dirac matrices
satisfying $\{\gamma_{M}^{\mu},\gamma_{M}^{\nu}\}=2\eta^{\mu\nu}$ with the Minkowski metric
$\eta^{\mu\nu}=\trm{diagonal}(+,-,-,-)$, while the matrices $\gamma^{\mu}(x)$ sastify the
corresponding anti-commutation relations with the arbitrary metric, i.e.
$\{\gamma^{\mu}(x),\gamma^{\nu}(x)\}=2\,g^{\mu\nu}(x)$. The inner product
$\big(\phi(x),\varphi(x) \big)_{\Sigma}$ is invariant under the choice of the
spacelike hypersurface $\Sigma$. For  the choice $t=\tau=0$ corresponding to
$n^{\mu} = (1,0,0,0)$ with $dS = dx dy dz$ in Minkowski coordinates, the inner product
reduces to the usual flat spacetime form $\int \phi^{\dagger}(x) \varphi(x) d^3x$.
In Rindler coordinates one can also choose to compute the inner product on the hypersurface $\tau=0$,
although it is often more convenient to perform the computation on the future horizon $H_+$, ($t=z$).
\bibitem{note2.5} The Minkowski annihilation operator for frequency $\omega_A$ is related to
the Rindler region I and region II operators of frequency $\omega_R$ more directly through an
intermediate set of modes called Unruh modes \cite{unruh,BD,carroll}. The Unruh modes
analytically extend the Rindler region I modes to region II, and the region II modes to region I as
discussed in the next paragraph. The Unruh modes are directly related to the Rindler region I and II
modes through a Bogoliubov transformations of the form \Eq{bogo_ops_a_bdag} and \Eq{bogo_ops_b_adag}
\cite{takagi,carroll}. Since the Unruh modes exist over all Minkowski space,
they share the same vacuum as the Minkowski annihilation operators. A Minkowski annihilation
operator of frequency $\omega_A$ (a frequency which Alice detects) can be written as an integral
(essentially a Fourier transform) over frequencies $\omega_R$ (that a Rindler observer detects)
involving only Unruh annihilation operators with phase factor coefficients (see\cite{takagi}).
The single mode approximation entails assuming that Rob's detector is sensitive to a single
particle mode frequency $\omega_R$ (an extremely narrow band detector) and that
the previously discussed frequency dependent phase factors
are highly peaked about $\omega_R \sim \omega_A$. We can then approximate the Minkowski
annihilation operator by a single Unruh annihilation operator,
which is then related to a Rindler particle annihilation operator
in region I and a Rindler anti-particle creation operator in region II through a Bogoliubov transformation
(see the discussion in the last paper in \cite{teleport}).
\bibitem{note3} Similar relations exist for the operators $c^{II}_{-k}$ and $d^{I}_{k}$ namely,
$c^{II}_{-k} |0_{-k}\rangle^+_{II} = 0$,
$d^{I}_k |0_k\rangle^-_{I} = 0$,
$c^{II\dagger}_{-k} |0_k\rangle^+_{II} = |1_{-k}\rangle^+_{II}$, and
$d^{I\dagger}_k |0_k\rangle^-_{I} = |1_k\rangle^-_{I}$.
However, these operators correspond to the mode $-k$, which we have ignored in the
single mode approximation.
\bibitem{note4} Let the intial unaccelerated state between Alice and Rob be the
tensor product of Minkowski states $\rho_A\otimes|\psi\rangle_R\langle \psi|$ where
$\rho_A$ is any state (possibly mixed) for Alice and $\psi$ is any pure state for Rob.
Since entanglement is invariant under local unitaries, we can always rotate $|\psi\rangle_R$ to
$|0_k\rangle^+$, the Minkowski vacuum for mode $k$ given in \Eq{vacM}, without changing
the initial entanglement. Then the reduced state for Rob is just
$|0_k\rangle^+\langle 0_k |$, which has maximal entanglement between I and II at infinite acceleration.
\end{thebibliography}
\end{document}